# A MULTI-HOP WEIGHTED CLUSTERING OF HOMOGENEOUS MANETS USING COMBINED CLOSENESS INDEX

T.N. Janakiraman[1] and A. Senthil Thilak[2]

[1, 2]Department of Mathematics, National Institute of Technology, Tiruchirapalli-620015, Tamil Nadu, India
[1]`janaki@nitt.edu, tnjraman2000@yahoo.com`
[2]`asthilak23@gmail.com`

## ABSTRACT

*In this paper, a new multi-hop weighted clustering procedure is proposed for homogeneous Mobile Ad hoc networks. The algorithm generates double star embedded non-overlapping cluster structures, where each cluster is managed by a leader node and a substitute for the leader node (in case of failure of leader node). The weight of a node is a linear combination of six different graph theoretic parameters which deal with the communication capability of a node both in terms of quality and quantity, the relative closeness relationship between network nodes and the maximum and average distance traversed by a node for effective communication. This paper deals with the design and analysis of the algorithm and some of the graph theoretic/structural properties of the clusters obtained are also discussed.*

## 1. AD HOC NETWORKS – A BRIEF REVIEW

An ad hoc wireless network is a collection of two or more devices (also termed as nodes) equipped with wireless communications and networking capability. Such devices/nodes can communicate either directly or through intermediate nodes depending on the availability of the nodes within or outside the radio range. An ad hoc network is self-organizing and adaptive, i.e, the already formed network can be de-formed on-the-fly without the need for any central administration. The nodes in an ad hoc network must be capable of identifying the connectivity with the neighbouring nodes, so as to allow communication and sharing of information and services. The nodes must perform routing and packet-forwarding functions. The topology changes continuously as the devices are not tied down to specific locations over time. Hence, the most important and challenging issues in a mobile ad hoc network are the mobile nature of the devices, scalability and constraints on resources such as limited bandwidth, limited and varying battery power, etc. Depending on the nature of devices, the uniformity in transmission range and network architecture, the network can either be homogeneous or heterogeneous. The network considered in this paper is a homogeneous where each node is assumed to have uniform transmission range.

## 2. SIGNIFICANCE OF CLUSTERING

A *cluster* is a subset of nodes of a network. *Clustering* is the process of partitioning a network into clusters and it is a way of making ad hoc networks more scalable. Scalability refers to the network's capability to facilitate efficient communication even in the presence of large number of network nodes. Cluster-based structures promote more efficient usage of resources in controlling large dynamic networks. With cluster-based control structures, the physical network is transformed into a virtual network of interconnected node clusters. Clustering can be done for different purposes, such as, clustering for transmission management, clustering for backbone management, clustering for routing efficiency etc, [1]. Each cluster has one or more controllers, such as leader nodes(also called as Masters or cluster-heads), Proxy nodes, Super-masters, gateways, etc. [2], acting on its behalf to make control decisions for cluster members and in some cases, to construct and distribute representations of cluster state for use outside of the cluster. The algorithm proposed in this paper is developed with the objective of facilitating routing functions by providing a hierarchical network organization and efficient sharing of resources and information.

In general, the process of clustering involves two phases, namely, cluster formation and cluster maintenance. Initially, the nodes are group together based on some principle to form the clusters. Then, as the nodes continuously move in different directions with different speeds, the existing links between the nodes also get changed and hence, the initially formed cluster structure cannot be retained for a longer period. So, it is necessary to go for the next phase, namely, cluster maintenance phase. Maintenance includes the procedure for modifying the cluster structure based on the movement of a cluster member outside an existing cluster boundary, battery drainage of cluster-heads, link failure, new link establishments, addition of a new node, node failure and so on.

## 3. PRIOR WORK

Several procedures are proposed and adopted for clustering of mobile ad hoc networks. Weight based clustering algorithms [4-7], Zone based clustering algorithms [8, 9], Dominating set based clustering [9, 10, 11] etc., are to name a few. In these clustering procedures, the clusters are formed based on different criteria and the algorithms are classified accordingly.

Based on whether a special node with specific features is required or not, the algorithms can be classified as cluster-head based and non-cluster-head based algorithms [11, 12]. Based on the hop distance between different pair of nodes in a cluster, they are classified as 1-hop clustering and multi-hop clustering procedures [11, 12]. Similarly, there exists a classification based on the objective of clustering, such as Dominating set based clustering, low maintenance clustering, mobility-aware clustering, energy efficient clustering, load-balancing clustering, combined-metrics based/weight based clustering [11]. This paper gives another different approach for clustering of such networks. As discussed in [13], the proposed algorithm is a weight based cum multi-hop clustering algorithm and is also an extension of 3hBAC [14] and LCC [15] clustering procedures. Hence, we give an overview of some of the algorithms coming under the two categories.

**LID Heuristic.** This is a cluster-head based, 1-hop, weight based clustering algorithm proposed by Baker and Ephremides [16, 17]. This chooses the nodes with lowest id among their neighbors as cluster-heads and their neighbors as cluster members. However, as it is biased to choose nodes with smaller ids as cluster-heads, such nodes suffer from battery drainage resulting in shorter life span. Also, because of having lowest id, a highly mobile node may be elected as a cluster-head, disturbing the stability of the network.

**HD Heuristic**. The highest degree (HD) heuristic proposed by Gerla et al. [18, 19], is again a cluster-head based, weight based, 1-hop clustering algorithm. This is similar to LID, except that node degree is used instead of node id. Node Ids are used to break ties in election. This algorithm doesn't restrict the number of nodes ideally handled by a cluster-head, leading to shorter life span. Also, this requires frequent re-clustering as the nodes are always under mobility.

**Least cluster change clustering (LCC)** [15]**.** This is an enhancement of LID and HD heuristics. To avoid frequent re-clustering occurring in LID & HD, the procedure is divided into two phases as in the proposed algorithm. The initial cluster formation is done based on lowest ids as in LID. Re-clustering is invoked only at instants where any two cluster-heads become adjacent or when a cluster member moves out of the reach of its cluster-head. Thus, LCC significantly improved stability but the second case for re-clustering shows that even the movement of a single node (a frequent happening in mobile networks) outside its cluster boundary will cause re-clustering.

**3-hop between adjacent clusterheads (3hBAC).** The 3hBAC clustering algorithm [14] is a 1-hop clustering algorithm which generates non-overlapping clusters. It assigns a new status, by name, cluster-guest for the network nodes apart from cluster-head and cluster member. Initially, the algorithm starts from the neighborhood of a node having lowest id. Then, the node possessing highest degree in the closed neighbor set of the above lowest id is elected as the initial cluster-head and its 1-hop neighbors are assigned the status of cluster members. After this, the subsequent cluster formation process runs parallely and election process is similar to HD heuristic. The cluster-guests are used to reduce the frequency of re-clustering in the maintenance phase.

**Weight-based clustering algorithms**. Several weight-based clustering algorithms are available in the literature [4-7], [14], [20], [21]. All these work similar to the above discussed 1-hop algorithms, except that each node is initially assigned a weight and the cluster-heads are elected based on these weights. The definition of node weight in each algorithm varies. Some are distributed algorithms [4], [20], [6] and some are non-distributed [5], [7], [14]. Each has its own merits and demerits.

**DSECA** [13]**.** The DSCEA is also a weight-based clustering which generates double star embedded non-overlapping structures, where the weight of each node is a linear combination of six parameters, namely, degree, node closeness index, mean hop distance, mean Euclidean distance and neighbour strength value. The algorithm proposed in this paper is a modified version of DSECA.

## 4. MODELLING ASSUMPTIONS

It is assumed that the network to be clustered is deployed by distributing the mobile nodes randomly in different positions on a terrain of size KxK.. Each node is assumed to have a uniform transmission range and the network under consideration is assumed to be homogeneous, unless otherwise specified. Those nodes within the transmission range of a particular node are identified as the 1-hop neighbors of that node. Each node identifies its 1-hop neighbors by transmitting Hello messages. The nodes are allowed to move randomly in different directions with varying velocity in the range $[0, V_{max}]$. To keep track of the changes in node positions due to mobility, the nodes send and receive Hello messages periodically at a predefined broadcast interval BI.

Each node computes its own weight and broadcasts a Weight_info() message containing its id, weight. Upon successful transmission and reception of Weight_info() messages by the entire set of nodes, each node maintains a weight table containing the weight information about all the other nodes in the network. Further, each node in the network has knowledge about the hop and Euclidean distance between itself and all the other nodes in the network. With these basic assumptions and information, the network nodes execute our proposed clustering procedure.

## 5. GRAPH PRELIMINARIES

A *graph G* is defined as an ordered pair *(V, E),* where *V* is a non-empty set of vertices/nodes and *E* denotes the set of edges/links between different pairs of nodes in *V.* Communication networks can in general be modeled using graphs. If any two nodes are within the transmission range of each other, then both can communicate with each other and are joined by a bidirectional link. The set of all nodes in the network is taken as the vertex (or node) set *V* of *G* and any two nodes are made adjacent (i.e., joined by a link) in *G*, if the corresponding two nodes can communicate with each other and the graph so obtained is called the *underlying graph or network graph or network topology*. Hence, the problem of "*Network Clustering*" can be viewed as a problem of "*Graph Partitioning*". Since each node is assumed to have uniform transmission range the underlying graph will always be an undirected graph.

If u and v are any two nodes in the network graph, then *d(u, v)* denotes the least number of hops to move from u to v and vice versa and is referred to as the *Hop-distance* between u and v and *ed(u, v)* denotes the Euclidean distance between u and v. Thus, in a homogeneous network, for a given transmission range r, two nodes u and v can communicate with each other only if they are at Euclidean distance less than or equal to r i.e., *ed(u, v)* ≤ *r*. Graph theoretically, two nodes u and v are joined by a link *e = (u, v)* or made adjacent in the network graph if their Euclidean distance is less than or equal to r, i.e., *ed(u, v)* ≤ *r*, else they are non-adjacent. The nodes u and v are called the *end nodes* of the link *e = (u, v)*. For a given node u, the neighbor set of u, denoted by *N(u),* is the set of those nodes which are within the transmission range of u, i.e, the set of those nodes which are 1-hop away from u and the cardinality of the set *N(u)* is defined as the *degree* of u and is denoted by *deg(u)*. The hop-distance between u and its farthest node in G is called the *eccentricity of u* in G and is denoted by *ecc(u)*, i.e., $ecc(u) = \max_{v \in V(G)} \{d(u,v)\}$. The average of the Hop-distances between u and each of the other nodes is defined to be *the mean-hop-distance of u* and is denoted by *MHD(u) i.e.,* $MHD(u) = \frac{1}{|V|}\left[ \sum_{v \in V(G)} d(u,v) \right]$. The average of the Euclidean distances between u and each of the other node is defined to be *the mean Euclidean distance of u* and is denoted by *MED(u) i.e.,* $MED(u) = \frac{1}{|V|}\left[ \sum_{v \in V(G)} ed(u,v) \right]$. The minimum and maximum eccentricities of *G* are defined respectively as *radius r(G)* and *diameter d(G)* of G .

A subset *S* of vertices of a graph *G* is said to be a *dominating set of G* if each vertex in *V-S* is adjacent to atleast one vertex in *S*. An *edge e = (u, v)* is said to *dominate an edge f* if either *f = (u, x)* or *f = (v, x)*, where *x* ∈ *V*. In other words, the edge *e = (u, v)* dominates an edge *f*, if *f* has atleast one of the vertices u or v as one of its end vertices [22]. An edge subset *E´* ⊆ *E* is an *efficient edge dominating set* for *G* if each edge in *E* is dominated by exactly one edge in *E´* [22].

The graph *G* which is rooted at a vertex say v, having n nodes $v_1$, $v_2$, … $v_n$, adjacent to v as shown in Figure 1, is called as the ***star graph*** and is denoted by $K_{1, n}$.

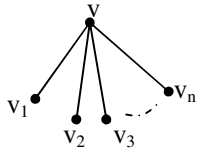

Figure 1. Star Graph

The graph obtained by joining the root vertices of the stars $K_{1, n}$ and $K_{1, m}$ by means of an edge as shown in Figure 2, is referred to as a ***double star*** or **(n, m)-*bi-star.***

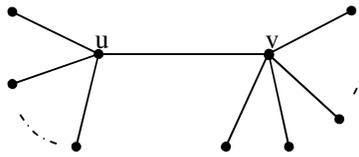

Figure 2. Double Star/(n, m)-bi-star

# 6. DEFINITION OF WEIGHT PARAMETERS

With the support of the idea generalized in [3], i.e., any meaningful parameter can be used as the weight to best exploit the network properties, here we use six different graph theoretic parameters for computing the weight of each node.

### 6.1. Closer-Hop set

Given a pair of nodes u and v in a graph *G*, the *closer-hop set of u relative to v,* is defined as the set of those nodes which are at a shorter hop distance with u compared to v and is denoted by *CHS(u|v)*, i.e., *CHS(u|v) = {w∈ V(G) : d(u, w) < d(v, w)}* and *$c_h(u|v)$ is the cardinality of the CHS(u|v)*. It is to be noted that $c_h(u|v)$ need not be equal to $c_h(v|u)$. In fact, $c_h(v|u) = N - c_h(u|v)$, where *N* denotes the total number of nodes in the network.

### 6.2. Closer-Euclidean set

Given a pair of nodes u and v in *G*, the *closer-euclidean set of u relative to v,* is defined as the set of those nodes which are at a shorter euclidean distance with u compared to v and is denoted by *CES(u|v)*, i.e., *CES(u|v) = { w in V(G) : ed(u, w) < ed(v, w)}* and *$c_{ed}(u|v)$ is the cardinality of the CES(u|v)*.

### 6.3. Hop-Closeness Index

Given two nodes u and v in a graph G, if $f_h(u, v) = c_h(u|v) - c_h(v|u)$, then the *hop-closeness index of u* denoted by $g_h(u)$, is defined as $g_h(u) = \sum_{v \in V(G)-u} f_h(u,v)$. For example, consider the graph in Figure 3. The Hop-closeness index of the node 1 is calculated as follows.

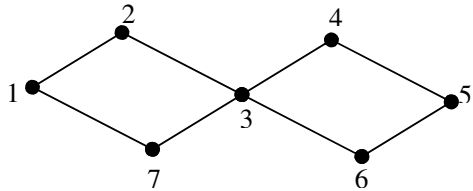

Figure 3. Graph for computing closeness index

$$g_h(1) = \sum_{i \in V(G)-1} f_h(1,i)$$

$$= f_h(1, 2) + f_h(1, 3) + f_h(1, 4) + f_h(1, 5) + f_h(1, 6) + f_h(1, 7)$$

$$= (-3) + (-1) + (0) + (-1) + (-3) + (-3)$$

$$= (-11)$$

### 6.4. Euclidean-Closeness Index

Given two nodes u and v in a graph G, if $f_{ed}(u, v) = c_{ed}(u|v) - c_{ed}(v|u)$, then the *euclidean-closeness index of u* denoted by $g_{ed}(u)$, is defined as $g_{ed}(u) = \sum_{v \in V(G)-u} f_{ed}(u,v)$. By knowing the (x, y) positions of each node in the network, the Euclidean-closeness index of each node can be computed in a similar fashion given in section 6.3.

If $g_h(u)$ (or $g_{ed}(u)$) is positive, then it indicates the positive relative closeness relationship, in the sense that, if for a node u, $g_h(u)$ (or $g_{ed}(u)$) is positive maximum, then it is more closer to all the nodes in the network compared to that of the other nodes. If $g_h(u)$ (or $g_{ed}(u)$) is negative maximum, it indicates that the node u, is highly deviated from all the other nodes in the network compared to that of the others. It gives a measure of the negative relative closeness relationship.

## 6.5. Combined-closeness index

The Combined-closeness index of a node u, denoted by $CCI(u)$ is defined to be the average of $g_h(u)$ and $g_{ed}(u)$. i.e, $CCI(u) = (g_h(u) + g_{ed}(u))/2$.

## 6.6. Categorization of neighbours of a node [13]

Depending on the Euclidean distance between the nodes, their signal strength varies. For a given node u (transmitting node), the nodes which are closer to u will receive stronger signals and those nodes which are far apart from u will get weaker signals. Based on this notion, the neighbors of a transmitting node are classified as follows:

    i.   Strong neighbours
    ii.  Medium neighbours
    iii. Weak neighbours

**Strong neighbour:** A node v is said to be a strong neighbour of a node u, if the Euclidean distance between u and v is less than or equal to r. i.e., $0 \le ed(u, v) \le r/2$.

**Medium neighbour:** A node v is said to be a medium neighbour of a node u, if $r/2 \le ed(u, v) \le 3r/4$.

**Weak neighbour:** A node v is said to be a weak neighbour of a node u, if $3r/4 \le ed(u, v) \le r$.

## 6.7. Neighbour Strength value

For any node u in the network, the neighbour strength value denoted by $NS(u)$ is defined to be $NS(u) = (m_1 + m_2/2 + m_3/4)K$, where K is any constant (a fixed threshold value) and $m_1, m_2, m_3$ denote respectively the number of strong, medium and weak neighbours of u. As explained in [13], for a node u with greater connectivity, its greater value is due to the contribution of all strong, weak and medium neighbours of u. But, if there exists another node v such that $deg(u) > deg(v)$ and $m_1(u) < m_1(v)$, $m_2(u) > m_2(v)$, $m_3(u) >>> m_3(v)$, then it is obvious that node u will be chosen because of having greater connectivity value. But, all its weak neighbours have greater tendency to move away from u. This affects the stability of u and hence the corresponding cluster, if u is chosen as a master/proxy. Hence, we use the parameter $NS(u)$ to determine the quality of the neighbours of a node and hence the quality of the links.

## 6.8. Node Weight

Since for real time applications, it is better to consider Euclidean distances rather than hop distances in some cases, and the hop distance cannot be ignored completely, in the proposed algorithm, instead of the node closeness index value used in the calculation of node weight in [13], we use the combined-closeness index value, by considering both Euclidean and hop distances. Thus, for any node u in the network, the weight of u, denoted by $W(u)$ is defined as follows.

$$W(u) = \alpha_1 \deg(u) + \alpha_2 CCI(u) + \alpha_3 (\frac{1}{ecc(u)}) + \alpha_4 (\frac{1}{MHD(u)}) + \alpha_5 (\frac{1}{MED(u)}) + \alpha_6 NS(u) \ ---- (1)$$

Here, the constants $\alpha_1, \alpha_2, \alpha_3, \alpha_4, \alpha_5$ and $\alpha_6$ are the weighing factors of the parameters under consideration and these may be chosen according to the application requirements. In the proposed algorithm, in order to give equal weightage to all the factors considered, we choose all the weighing factors as (1/n), where n = number of parameters considered. Here, n = 6.

## 7. STATUS OF THE NODES IN A NETWORK GRAPH

In the proposed algorithm, each node in the network is assigned one of the following status:

* **Master** – A node which is responsible for coordinating network activities and also responsible for inter and intra cluster communication
* **Proxy** – A node adjacent to a master node which plays the role of a master in case of any failure of the master.
* **Slaves** – Neighbors of Master nodes and/or Proxy nodes
* **Type I Hidden Master** – A neighbor node of a Proxy having greater weight than proxy.
* **Type II Hidden Master** – A node with greater weight and eligible for Master/Proxy selection, but not included in cluster formation because of not satisfying distance property and also not adjacent to any Proxy node.
* A node which is neither a slave nor a Master/Proxy.

It is to be noted that a node which was a type II hidden master at some instant may become a type I hidden master at a later instant.

## 8. BASIS OF OUR ALGORITHM

In all cluster-head based algorithms, a special node called a cluster-head plays the key role in communication and controlling operations. These cluster-heads are chosen based on different criteria like mobility, battery power, connectivity and so on. Though, a special care is taken in these algorithms to ensure that the cluster-heads are less dynamic, the excessive battery drainage of a cluster-head or the movement of a cluster-head away from its cluster members require scattering of the nodes in the cluster structure and re-affiliation of all the nodes in that cluster.

To overcome this problem, in the proposed algorithm, in addition to the cluster-heads (referred to as Masters in our algorithm), we choose another node called Proxy, to act as a substitute for the cluster-head/master, when the master gives up its role and also to share the load of a cluster-head. In our algorithm, each node is assigned a weight based on different criteria. The weight of a node is a linear combination of six different parameters as in (1). The algorithm concentrates on maximum weighted node and the weight is maximum if the parameters $deg(u)$, $g(u)$, $NS(u)$ are maximum and $ecc(u)$, $MHD(u)$ and $MED(u)$ are all minimum. The following characteristics are considered while choosing the parameters.

1. The factor $deg(u)$ denotes the number of nodes that can interact with u or linked to u, which is otherwise stated as the connectivity of the node. By choosing a node u with $deg(u)$ to be maximum, we are trying to choose a node having higher connectivity. This will minimize the number of clusters generated.
2. The metric, neighbour strength value, denoted by $NS(u)$ gives the quality of the links existing between a node and its neighbors. By choosing $deg(u)$ and $NS(u)$ to be simultaneously maximum, we give preference to a node having good quantity and quality of neighbours/links.
3. The parameter $CCI(u)$ gives a measure of the relative closeness relationship between u and the other nodes in the network, both in terms of hop and Euclidean distances. By choosing $CCI(u)$ to be maximum, we are concentrating on the node having greater affinity towards the network.
4. By choosing a node with minimum $ecc(u)$, we concentrate on a node which is capable of communicating with all the other nodes in least number of hops compared to others.
5. By selecting a node with minimum MHD and MED, we choose a node for which the average time taken to successfully transmit the messages (measured both in terms of number of hops and Euclidean distance) among/to the nodes in the network is much lesser.

## 9. OBJECTIVES OF THE ALGORITHM

The algorithm discussed in this paper is designed with the following objectives.

1. The network nodes are partitioned into different groups of various sizes to form a hierarchical organization of the network.
2. The cluster formation and maintenance overheads should be minimized.
3. The clusters generated must be stable as long as possible.
4. The leader nodes should not be overloaded. Here, it is distributed between the master and proxy nodes.
5. Re-affiliations should be minimized.
6. Re-clustering should be avoided as much as possible. At times of necessity, re-affiliations are allowed instead of re-clustering to reduce the cost of cluster maintenance.
7. The algorithm should overcome the problem of scalability.
8. The generated clusters should facilitate hierarchical routing.

## 10. PROPOSED ALGORITHM – MODIFIED DSECA (M_DSECA)

The proposed algorithm is a modified version of DSECA given in [13]. The algorithm is an extension of the 3hBAC clustering algorithm [14] and the weight-based clustering algorithms. As in LCC, 3hBAC and other clustering algorithms, the proposed clustering procedure also involves two phases, namely, cluster formation phase and cluster maintenance phase.

### 10.1. Notations used in M_DSECA

* The tuple **(m, p)** denotes *a master and its corresponding proxy pair.*

* **(M, P)** denotes *the set of all (m, p) such that m is a master and p is its corresponding proxy pair.*

* **hm-I** denotes *a node which is a hidden master of type I.*

* **HM-I** denotes *the set of those nodes which are hidden masters of type I.*

* **hm-II** denotes *a node which is a hidden master of type II.*

* **HM-II** denotes *the set of those nodes which are hidden masters of type II.*

* **N(u)** denotes *the set of all 1-hop neighbours of u.*

* **N'(u)** denotes *the set of those neighbours of u having greater weight than u.*

* **N''(u)** denotes *the set of those neighbours of u which are not Master/Proxy nodes and also having lesser weight than that of u.*

* $N_m(u)$ denotes *the set of those neighbours of u which are adjacent to some Master node.*

### 10.2. Cluster set up phase

**M_DSEC(*G*).**

In the cluster set up phase, initially, all the nodes are grouped into some clusters.

***Initial (Master, Proxy) election.*** Among all the nodes in the network, choose a node having maximum weight. It is designated as a Master. Next, among all its neighbors, the one with greater weight is chosen and it is designated as a Proxy. Then the initial cluster is formed with the chosen Master, Proxy and their neighbors. Since this structure will embed in itself a double star, the algorithm is referred to as a double star embedded clustering algorithm.

***Second and subsequent (Master, Proxy) election.*** For the subsequent (Master, Proxy) elections, we impose an additional condition on the hop distance between different (m, p) pairs to generate non-overlapping clusters. Here, as in 3hBAC, we impose the condition that all the (m, p) pairs should be atleast 3 hops away from each other. Nodes which are already grouped into some clusters are excluded in the future cluster formation processes. Among the remaining pool of nodes, choose the one with higher weight. Next,

(i) Check whether the newly chosen node is exactly 3-hop away from atleast one of the previously elected Masters (or Proxies) and atleast 3-hop from the corresponding Proxy (or Master)

(ii) The newly elected node should be at distance atleast 3-hop from rest of the (m, p) pairs.

If the above chosen higher weight node satisfies these two conditions, it can be designated as a Master.

To choose the corresponding Proxy, among the neighbours of above chosen Master, find the one with higher weight and at distance atleast 3-hop from each of the previously elected (m, p) pairs. Then, obtain a new cluster with this chosen (m, p) pair and their neighbours. Repeat this procedure until all the nodes are exhausted. The nodes which are not grouped into any cluster and the set HM-I are collected separately and termed as "Critical nodes". The set of all nodes grouped into some clusters is denoted by S and the set of critical nodes is denoted by C. Hence, after the cluster set up phase, the sets S and C are obtained as output. The pseudo code for the above process is given below.

### 10.2.1. Main Procedure

**M_DSEC(G)**

1. Randomly generate the required node positions of all the nodes in the network.
2. **for** each node u∈ N, compute N(u)
3. Compute the Euclidean distance matrix and Hop distance matrix.
4. S=3-hop-M_DSEC(N)  /*Calling procedure to form a set with maximum possible 3-hop perfect double star embedded clusters*/
5. C = N/S $\cup$ CR        // C forms the Critical node set
6. **If**(C == $\phi$ )
7.    **then**
        **Print** "Perfect clustering…" & goto step 15
8.    **else {**
9.       $S_A$ = adjusted_M_DSEC(S, C)
10.      $C_A$ = N\$S_A$
11.      **Print** "Refined Clustering…"
12.      **Return** $S_A$, $C_A$
13.      **Exit**
14. **}Endif**
15. **Return S, C**
16. **Exit**

### 10.2.2. Formation of perfect 3-hop modified Double star embedded clusters

**3-hop-M_DSEC(N)**

1.    **for** each vertex u∈ N, compute W(u)
2.    S$\leftarrow \phi$     //Union of all double star embedded clusters

3.     CR $\leftarrow \phi$   //Union of hidden masters of type I

4.     j $\leftarrow$ 1

5.     Extract a node, say x, from N such that W(x) is maximum. (In case of a tie, choose the one with higher NS value)

6.     Find N(x)//Nodes within one hop from x

7.     From N(x), extract a node with maximum weight. Label it as y.

8.     Find N(y) and N'(y) = set of those neighbors of y having greater weight than y

9.     $C_j \leftarrow \{x, y\} \cup N(x) \cup N(y)$   /* Initial cluster formation, x acts as master, y acts as proxy and its neighbors are slaves */

10.   Master[ j]$\leftarrow$x, Proxy[ j]$\leftarrow$y, HM[j]$\leftarrow$ N'(y)

11.   S$\leftarrow$S $\cup$ $C_j$ //Updation of double star embedded clusters

12.   CR = CR $\cup$ HM[j]

13.   P=$\phi$   /*Set of those nodes with higher weight but not eligible for Master because of not satisfying distance property */

14.   **do{**

15.   Extract a node, say x, from N\[S $\cup$ P] such that W(x) is maximum. (In case of a tie, choose the one with higher NS value). Label the newly chosen node as z.

16.   **If**((d(z, Master[i])==3) && d(z, Proxy[i]$\geq$3)) ||

        (d(z, Proxy[i]==3) && d(z, Master[i]$\geq$3)), for some 1$\leq$i$\leq$j) {

17.   **If**((d(z, Master[k])$\geq$3&&d(z, Proxy[k])$\geq$3),

        for all k$\neq$ i and 1$\leq$ i, k$\leq$j) {

18.        j$\leftarrow$j+1

19.        x $\leftarrow$ z

20.        Goto step 6

21.        }

22.   **else** {

23.       P = P $\cup$ z

24.       Goto step 14

25.   }**Endif**

26.   **else** {

27.       P = P $\cup$ z

28.       Goto step 14

29.   }**Endif**

30.   }**while**(N\[S $\cup$ P]$\neq$ $\phi$ )

31.   **Return** S, CR

32.   **Exit**

## 10.3. Cluster Maintenance phase – Treatment of critical nodes generated by M_DSECA

### 10.3.1. Nature of Critical nodes

A node in the critical node set C generated after implementing M_DSECA may be of any one of the following categories:
(i)  A Hidden Master of type I.
(ii)  A Hidden Master of type II.
(iii) A Node neglected in cluster formation because of lesser weight.

### 10.3.2. Neighbors of critical nodes

Let u be a critical node and v be a neighbor of u. Then the following cases may arise:

*Case i: v is another critical node*
<u>Subcase i:</u> v is a hm-I.
Then v will have an adjacent Proxy node such that w(v) is greater than that of the proxy node. In this case, the set N''(v) will be used to form adjusted clusters.
<u>Subcase i:</u> v is not a hm-I.
In this case, the set N''(v)\N$_m$(v) will be used for adjusted cluster formation.

*Case ii: v is a slave node (an existing cluster member)*
In this case, v will be used for adjusted cluster formation provided v is not adjacent to any existing Master. In such a case, we consider all the neighbors of v having lesser weight than v except the neighbors which are Proxy nodes.
If v is adjacent to some master, then it will not be used for adjusted cluster formation.

### 10.3.3. Formation of adjusted Double star embedded clusters

In the formation of adjusted clusters, we try to form clusters of these critical nodes either among themselves or by extracting nodes from existing clusters and regroup them with critical nodes to form better clusters. The below adjusted clustering procedure is invoked to minimize the number of critical nodes.

**adjusted_M_DSEC(S, C)**

From C, extract a node with maximum weight. Let it be c. Then any one of the following cases arises. Here, we form the adjusted clusters depending on the nature of the critical nodes by considering only restricted neighbours as explained in section 10.3.2.

*Case i: c is a hm-I.* Then, c will have an adjacent Proxy, say p. From N(c)\{p}, choose a node, say c', having greater weight.

*Subcase i: c' is another critical node.* If c' is a hm-I, then find N''(c') and form the new adjusted double star embedded cluster with {c, c'} $\cup$ (N(c)\{p}) $\cup$ N''(c'). The node c acts as the Master and c' as the Proxy of the new adjusted cluster. Otherwise, the set {c, c'} $\cup$ (N(c)\{p}) $\cup$ N(c') will form the new adjusted double star embedded cluster with c as Master and c' as Proxy.

*Subcase ii: c' is a slave node.* In this case, c' may be adjacent to some Master/Proxy of existing clusters. As explained in 10.3.2., if c' is adjacent to any master, then it will not be used for adjusted cluster formation. If not, then we obtain a new adjusted cluster with {c, c'} $\cup$ (N(c)\{p}) $\cup$ N''(c').

*Case ii: c is not a hm-I.* In this case, from N''(c)\N$_m$(c), choose a node, say c' having greater weight, then form a new adjusted cluster, with c, c', N''(c)\N$_m$(c) and N''(c')\N$_m$(c').

Repeat this procedure until either all the nodes are exhausted or no such selection can be performed further. If there is any node still left uncovered after completing this procedure, it will become a Master on its own.

Further, as the position of nodes may change frequently due to mobility, each (m, p) pair should periodically update its neighbor list so that if any slave node moves outside its cluster boundary, it can attach itself to its neighboring cluster by passing find_CH messages to all (m, p) pairs. If it receives an acknowledgment from some Master/Proxy, it will join that cluster. In case getting an acknowledgement from two or more nodes, the slave chooses the one with higher weight.

## 11. AN ILLUSTRATION

The above given procedure is explained with the following network graph. Consider the network graph shown in Figure 4., consisting of 23 nodes. The (x, y) positions of the nodes in the network are randomly generated and the graph is plotted with those positions. The weight of each node is computed using formula 1. Appendix gives a detailed description of the computation of weights of the nodes in the below network graph. The values in the parentheses denote the node id and weight values of the respective nodes. Here, the value of NS(u) is computed by arbitrarily considering some of the neighbours as strong, some as weak and some as medium neighbours and taking the threshold value K=100.

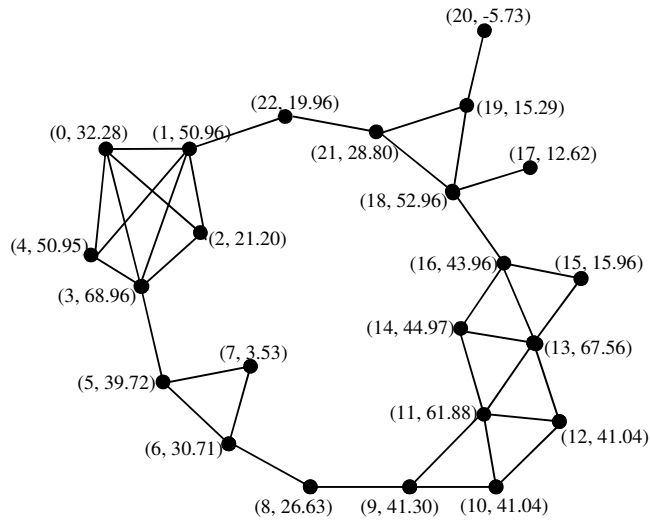

Figure 4. An Example Network Graph (G)

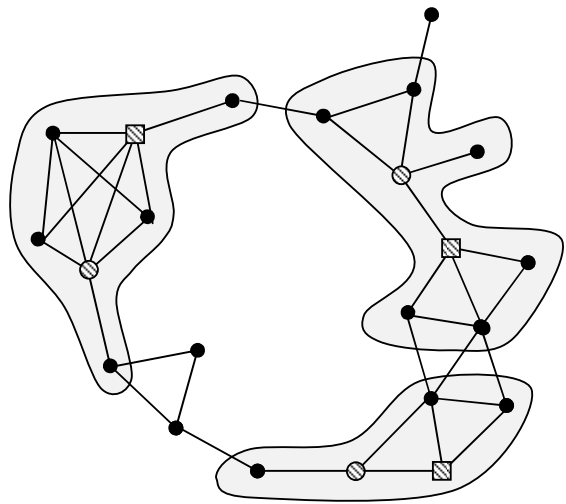

Figure 5. Clusters generated after executing **M_DSEC(G)**

*(Here, the striped circles denote Masters, Striped squares denote Proxies and Shaded circles denote slaves)*

**Clusters generated after executing M_DSEC(G):**

$C_1 = \{(\mathbf{3}, \mathbf{1}), 0, 2, 4, 5, 22\}$
$C_2 = \{(\mathbf{18}, \mathbf{16}), 13, 14, 15, 17, 19, 21\}$
$C_3 = \{(\mathbf{9}, \mathbf{10}), 8, 11, 12\}$

After executing **M_DSEC(G),** the adjustment procedure **adjusted_M_DSEC(S, C)** explained in section 10.3.3. is executed with S = $C_1 \cup C_2 \cup C_3$ and HM-I = {11, 13, 14}, HM-II = {11, 13}, C = HM-I $\cup$ Set of nodes left unclustered = {11, 13, 14, 6, 7, 20}. Among the nodes in C, the node 13 possesses highest weight. Hence, it becomes an eligible master for adjusted cluster formation. Now, node 13 is a hm-I. Therefore, it has an adjacent proxy, i.e., node 16. Hence, by looking into N(13)\{16}, we choose a node with higher weight, which node 11. Thus, by using case (i) of section 10.3.3., we get a new adjusted cluster $C_1$' = {(**13, 11**), 12, 14, 15}. At the same time cluster $C_2$ gets changed as $C_2$ = {(**18, 16**), 17, 19, 21}. Then continuing with the remaining set of critical nodes, i.e., {6, 7, 20}, we get another new adjusted cluster $C_2$' = {6, 7}. The node 20 is still left uncovered. So, it is declared as a master on its own. Thus, the adjusted clusters obtained finally will be as shown in Figure 6. It can be seen from Figure 5 and Figure 6 that after implementing the adjustment procedure, not only the critical nodes are grouped into some clusters, but also the already generated clusters get adjusted automatically so that the load is well balanced. Hence, the algorithm generates optimum load balancing clusters.

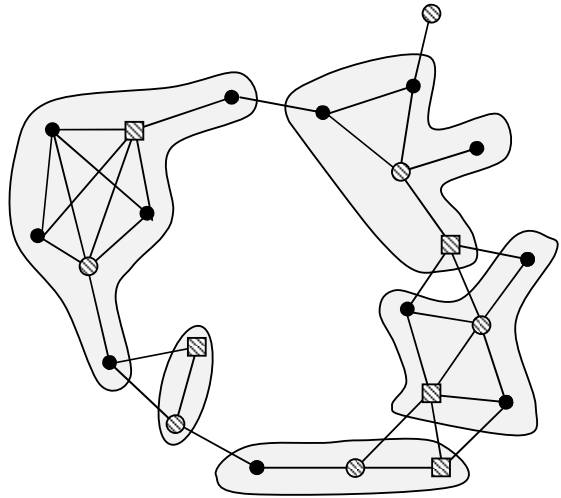

Figure 6. Adjusted Clusters formed after executing **adjusted_M_DSEC**

## 12. CATEGORIZATION OF M_DSE CLUSTERING

Any clustering which yields no critical nodes after initial cluster formation is said to be a ***perfect clustering.*** The one which yields some critical nodes but the number can be reduced to zero after the execution of adjustment procedure given in 9.2.3. is said to be a ***fairly-perfect clustering*** scheme and the one in which the number of critical nodes cannot be reduced to zero even after implementing the adjustment procedure is said to be an ***imperfect clustering.***

## 13. PROPERTIES OF THE CLUSTER STRUCTURES

In general, to meet the requirements of the ad hoc networks, a clustering algorithm is required to partition the nodes of the network so that three ad hoc clustering properties are satisfied. (1) Dominance Property, (2) Independence Property, (3) Guaranteed good service by the leader nodes.[3, 21]. It can be seen that the proposed algorithm also satisfies the above properties. i.e.,

1. Every ordinary node (a node which is neither a master nor a proxy) affiliates with a leader node (Master/Proxy) (dominance property).

2. As per the proposed algorithm, the Master nodes are always maintained to have higher weight than the rest of the nodes in that cluster.

3. Every slave node is at most $d$ hops away from its (Master, Proxy)-pairs, where $d = 2$.

4. No two Master nodes are adjacent (guarantees well scattered clusters).

The following are some of the other graph theoretic/structural properties observed in the cluster structures obtained using the proposed algorithm.

**Property 1.** Each cluster is of diameter atmost 3.

**Property 2.** Each double star embedded subgraph has a dominating edge

**Property 3.** After finishing the execution of both M_DSEC and the adjustment procedure, each vertex lies in exactly one cluster, as each slave node is affiliated with exactly one (Master, Proxy)-pair, whereas each critical node is declared itself as a leader node.

**Property 4.** If the resultant clustering is a *perfect clustering,* then the set of all (Master, Proxy)-pairs will form an *efficient edge dominating set* and the total number of clusters obtained in such a case will be equal to the domination number of the line graph of the underlying network graph.

## 14. CONCLUSION AND FUTURE WORK

The proposed algorithm yields a cluster structure, where the clusters are managed by Master nodes. In case of any failure of Master nodes, the cluster is not disturbed and the functions are handed over to an alternative which behave in a similar way to Master nodes (Perhaps with less efficiency than Masters but better than ordinary nodes). In order to better suite practical constraints, we have included the Euclidean-closeness measures in addition with the hop-closeness measure used in [13]. This enables us to increase the life time of the network. Further, since the clusters are managed by the Masters as well as by the Proxy nodes at times of necessity, the load is well balanced. The event of re-clustering can be avoided as long as possible. The algorithm is being implemented in NS2 and the expected better performance of the algorithm will be guaranteed on comparison of this with the other existing algorithms.

## REFERENCES

[1]    C.E. Perkins, "Ad hoc Networking", Addison Wesley, Pearson Education, Inc. And Dorling Kindersley Publishing, Inc., India © 2001.

[2]    L. Ramachandran, M. Kapoor, A. Sarkar and A. Aggarwal, "Clustering Algorithms for wireless ad hoc networks", Proc. of 4th International Workshop on Discrete algorithms and methods for mobile computing and communications, Boston, MA, August 2000, pp. 54-63.

[3]    S. Basagni, "Distributed Clustering for Ad hoc networks", Proc. of ISPAM'99 International Symposium on parallel architectures, algorithms and networks, pp. 310-315, 1999.

[4]  M. Chatterjee, S.K. Das and D. Turgut, "A Weight based distributed clustering algorithm for MANET", V.K. Prasanna, et. al (eds.) HiPC 2000, LNCS, vol. 1970, pp. 511-521, Springer, Heidelberg (2000).

[5]  M. Chatterjee, S.K. Das and D. Turgut, "WCA: A Weighted Clustering Algorithm for Mobile Ad Hoc Networks", Cluster Computing, vol. 5, pp. 193-204, Kluwer Academic Publishers, The Netherlands, 2002.

[6]  W. Choi and M. Woo, "A Distributed Weighted Clustering algorithm for mobile ad hoc networks", Proc. of AICT/ICIW 2006, IEEE, Los Alamitos, 2006.

[7]  W. Yang and G. Zhang, "A Weight-based clustering algorithm for mobile ad hoc networks", Proc. of 3rd International Conference on Wireless and mobile communications, 2007.

[8]  I.Y. Kim, Y.S. Kim and K.C. Kim, "Zone-based clustering for intrusion detection architecture in ad hoc networks", Management of Convergence networks and services, APNOMS 2006 Proceedings, LNCS, vol. 4238, pp. 253-262, Springer, Heidelberg, 2006.

[9]  Y.P. Chen, A.L. Liestman and J. Liu, "Clustering Algorithms for ad hoc wireless networks", in Ad hoc and Sensor Networks, Y. Pan and Y. Xiao (eds.), Nova Science Publishers, pp. 1-16, 2004.

[10]  K. Erciyes, O. Dagdeviren, D. Cokuslu and D. Ozsoyeller, "Graph Theoretic clustering algorithms in mobile ad hoc networks and wireless sensor networks Survey", Appl. Comput. Math. Vol. 6, No. 2, pp. 162-180, 2007.

[11]  J.Y. Yu and P.H.J. Chong, "A survey of clustering schemes for mobile ad hoc networks", First Quarter, Vol. 7, No. 1, pp. 32-47, 2005.

[12]  S.J. Francis, E.B. Rajsingh, "Performance analysis of clustering protocols in mobile ad hoc networks", J. Computer Science, Vol. 4, No. 3, pp. 192-204, 2008.

[13]  T.N. Janakiraman and A.S. Thilak, "A Weight based double star embedded clustering of homogeneous mobile ad hoc networks using graph theory", Advances in Networks and Communications, N. Meghanathan et al. (eds.), CCIS, Vol. 132, Part – II, pp. 329-339, Proc. of CCSIT 2011, Springer, Heidelberg, 2011.

[14]  J.Y. Yu and P.H.J. Chong, "3hBAC (3-hop Between Adjacent cluster heads: a novel non-overlapping clustering algorithm for mobile ad hoc networks", Proc. of IEEE Pacrim 2003, Vol. 1, pp. 318-321, 2003.

[15]  C. Chiang, "Routing in clustered multihop, mobile wireless networks with fading channel", Proc. of IEEE SICON '97, 1997.

[16]  A.A. Abbasi, M.I. Buhari and M.A. Badhusha, "Clustering Heuristics in wireless networks: A survey", Proc. 20th European Conference on Modelling and Simulation, 2006.

[17]  D.J. Baker and A. Epremides, "A distributed algorithm for organizing mobile radio telecommunication networks", Proc. 2nd International conference on distributed computer systems, pp. 476-483, IEEE Press, France, 1981.

[18]  M. Gerla and J.T.C. Tsai, "Multi-cluster, mobile, multimedia radio network", Wireless networks, vol. 1, No. 3, pp. 255-265, 1995.

[19]  A.K. Parekh, "Selecting routers in ad hoc wireless networks", Proc. SB/IEEE International Telecommunications Symposium, IEEE, Los Alamitos, 1994.

[20]  P. Basu, N. Khan, T.D.C. Little. "A mobility based metric for clustering in mobile ad hoc networks", Proc. IEEE ICDCS, Phoenix, Arizona, USA, pp. 413-418, 2001.

[21]  E.R. Inn and W.K.G. Seah, "Performance analysis of mobility-based d-hop (MobDHop) clustering algorithm for mobile ad hoc networks", Computer Networks, vol. 50, 3339-3375, 2006.

[22]  T.W. Haynes, S.T. Hedetniemi and P.J. Slater, "Fundamentals of Domination in Graphs", Marcel Dekker, Inc., New York.

**Authors**

**T.N. Janakiraman** is currently Associate Professor of Department of Mathematics, National Institute of Technology, Tiruchirapalli, India. He completed his undergraduate Studies at Madras University, India in 1980 and completed his Post graduation at National College, Trichy, India in 1983. He did his Ph.D. in Mathematics (Graph Theory and its applications) at Madras University with a UGC sponsored research fellowship and received his doctoral degree in the year 1991. He was a Postdoctoral Research associate for 1 year (1993-1994) in Madras University under the He has two sponsored research projects to his credit and published around 70 papers in refereed National/International journals. His research interests include Pure Graph Theory, Applications of Graph Theory to Fault tolerant networks, Central location problems, Clustering of wired & wireless ad hoc networks, Clustering of cellular and flexible manufacturing models, Image processing, Graph coding and Graph Algorithms.

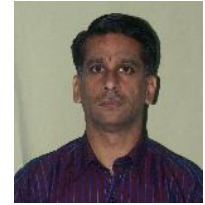

**A. Senthil Thilak** is currently a research Scholar of Department of Mathematics, National Institute of Technology, Tiruchirapalli, India. She received her Master's degree in Mathematics and Master of Philosophy in Mathematics from Seethalakshmi Ramaswami College, Tiruchirapalli, India. She has completed Post Graduate Diploma in Computer Applications in Bharathidasan University, Tiruchirapalli, India. She has published three papers in refereed National/International Journals. Her main research interests include Pure Graph Theory, Algorithmic Graph Theory and applications of graph theory to wireless ad hoc networks.

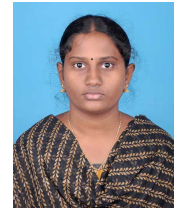

# APPENDIX

## Table 1. DISTANCE MATRIX

| Node (u) | 0 | 1 | 2 | 3 | 4 | 5 | 6 | 7 | 8 | 9 | 10 | 11 | 12 | 13 | 14 | 15 | 16 | 17 | 18 | 19 | 20 | 21 | 22 | Deg(u) | ecc(u) | 1/ecc(u) | 1/MHD(u) | g_h(u) |
|---|---|---|---|---|---|---|---|---|---|---|---|---|---|---|---|---|---|---|---|---|---|---|---|---|---|---|---|---|
| 0 | 0 | 1 | 1 | 1 | 1 | 2 | 3 | 3 | 4 | 5 | 6 | 6 | 7 | 6 | 6 | 6 | 5 | 4 | 4 | 5 | 3 | 2 | 4 | 7 | 0.14 | 0.27 | -47 |
| 1 | 1 | 0 | 1 | 1 | 1 | 2 | 3 | 3 | 4 | 5 | 6 | 6 | 6 | 5 | 5 | 5 | 4 | 4 | 3 | 3 | 4 | 2 | 1 | 5 | 6 | 0.17 | 0.31 | 46 |
| 2 | 1 | 1 | 0 | 1 | 2 | 2 | 3 | 3 | 4 | 5 | 6 | 6 | 7 | 6 | 6 | 6 | 5 | 5 | 4 | 4 | 5 | 3 | 2 | 3 | 7 | 0.14 | 0.26 | -54 |
| 3 | 1 | 1 | 1 | 0 | 1 | 1 | 2 | 2 | 3 | 4 | 5 | 5 | 6 | 6 | 6 | 5 | 5 | 4 | 4 | 5 | 3 | 2 | 5 | 6 | 0.17 | 0.29 | 18 |
| 4 | 1 | 1 | 2 | 1 | 0 | 2 | 3 | 3 | 4 | 5 | 5 | 6 | 7 | 6 | 6 | 6 | 5 | 5 | 4 | 5 | 3 | 2 | 3 | 7 | 0.14 | 0.27 | -44 |
| 5 | 2 | 2 | 2 | 1 | 2 | 0 | 1 | 2 | 3 | 4 | 4 | 5 | 5 | 5 | 6 | 6 | 5 | 5 | 6 | 4 | 3 | 3 | 6 | 7 | 0.14 | 0.29 | -10 |
| 6 | 3 | 3 | 3 | 2 | 3 | 1 | 0 | 1 | 2 | 3 | 3 | 4 | 4 | 4 | 5 | 5 | 7 | 6 | 6 | 7 | 5 | 4 | 3 | 7 | 0.14 | 0.28 | -14 |
| 7 | 3 | 3 | 3 | 2 | 3 | 1 | 1 | 0 | 2 | 3 | 4 | 4 | 5 | 6 | 6 | 7 | 6 | 7 | 5 | 4 | 2 | 7 | 0.14 | 0.25 | -77 |
| 8 | 4 | 4 | 4 | 3 | 4 | 2 | 1 | 2 | 0 | 1 | 2 | 2 | 3 | 3 | 3 | 4 | 4 | 6 | 5 | 6 | 7 | 6 | 5 | 2 | 7 | 0.14 | 0.28 | -14 |
| 9 | 5 | 5 | 5 | 4 | 5 | 3 | 2 | 3 | 1 | 0 | 1 | 1 | 2 | 2 | 2 | 3 | 3 | 5 | 4 | 5 | 6 | 5 | 6 | 3 | 6 | 0.17 | 0.29 | 10 |
| 10 | 6 | 6 | 6 | 5 | 5 | 4 | 3 | 4 | 2 | 1 | 0 | 1 | 1 | 2 | 2 | 3 | 3 | 5 | 4 | 5 | 6 | 5 | 6 | 3 | 6 | 0.17 | 0.27 | -43 |
| 11 | 6 | 6 | 6 | 5 | 6 | 4 | 3 | 4 | 2 | 1 | 1 | 0 | 1 | 1 | 1 | 2 | 2 | 4 | 3 | 4 | 5 | 4 | 5 | 5 | 6 | 0.17 | 0.30 | 33 |
| 12 | 7 | 6 | 7 | 6 | 7 | 5 | 4 | 5 | 3 | 2 | 1 | 1 | 0 | 1 | 2 | 2 | 2 | 4 | 3 | 4 | 5 | 4 | 5 | 3 | 7 | 0.14 | 0.27 | -43 |
| 13 | 6 | 5 | 6 | 6 | 6 | 5 | 4 | 5 | 3 | 2 | 2 | 1 | 1 | 0 | 1 | 1 | 1 | 3 | 2 | 3 | 4 | 3 | 5 | 5 | 6 | 0.17 | 0.31 | 46 |
| 14 | 6 | 5 | 6 | 6 | 6 | 6 | 5 | 6 | 3 | 2 | 2 | 1 | 2 | 1 | 0 | 2 | 1 | 3 | 2 | 3 | 4 | 3 | 4 | 3 | 6 | 0.17 | 0.30 | 39 |
| 15 | 6 | 5 | 6 | 6 | 6 | 6 | 5 | 6 | 4 | 3 | 3 | 2 | 2 | 1 | 2 | 0 | 1 | 3 | 2 | 3 | 4 | 3 | 4 | 2 | 6 | 0.17 | 0.28 | -20 |
| 16 | 5 | 4 | 5 | 5 | 5 | 6 | 5 | 6 | 4 | 3 | 3 | 2 | 2 | 1 | 1 | 1 | 0 | 2 | 1 | 2 | 3 | 2 | 3 | 4 | 6 | 0.17 | 0.32 | 84 |
| 17 | 5 | 4 | 5 | 5 | 5 | 6 | 7 | 7 | 6 | 5 | 5 | 4 | 4 | 3 | 3 | 3 | 2 | 0 | 1 | 2 | 3 | 2 | 3 | 1 | 7 | 0.14 | 0.26 | -52 |
| 18 | 4 | 3 | 4 | 4 | 4 | 5 | 6 | 6 | 5 | 4 | 4 | 3 | 3 | 2 | 2 | 2 | 1 | 1 | 0 | 1 | 2 | 1 | 2 | 4 | 6 | 0.17 | 0.33 | 105 |
| 19 | 5 | 4 | 5 | 5 | 5 | 4 | 5 | 6 | 6 | 5 | 5 | 4 | 4 | 3 | 3 | 3 | 2 | 2 | 1 | 0 | 1 | 1 | 2 | 3 | 6 | 0.17 | 0.29 | 34 |
| 20 | 5 | 4 | 5 | 5 | 5 | 6 | 7 | 7 | 7 | 6 | 6 | 5 | 5 | 4 | 4 | 4 | 3 | 3 | 2 | 1 | 0 | 2 | 3 | 1 | 7 | 0.14 | 0.23 | -128 |
| 21 | 3 | 2 | 3 | 3 | 3 | 4 | 5 | 5 | 6 | 5 | 5 | 4 | 4 | 3 | 3 | 3 | 2 | 2 | 1 | 1 | 2 | 0 | 1 | 3 | 6 | 0.17 | 0.33 | 89 |
| 22 | 2 | 1 | 2 | 2 | 2 | 3 | 4 | 4 | 5 | 6 | 6 | 5 | 5 | 5 | 4 | 4 | 3 | 3 | 2 | 2 | 3 | 1 | 0 | 2 | 6 | 0.17 | 0.31 | 42 |

Table 2. EUCLIDEAN DISTANCE MATRIX

| u | 0 | 1 | 2 | 3 | 4 | 5 | 6 | 7 | 8 | 9 | 10 | 11 | 12 | 13 | 14 | 15 | 16 | 17 | 18 | 19 | 20 | 21 | 22 | 1/MED(u) | E_ref(u) |
|---|---|---|---|---|---|---|---|---|---|---|----|----|----|----|----|----|----|----|----|----|----|----|----|----------|----------|
| 0 | 0.00 | 0.32 | 0.64 | 0.89 | 2.31 | 3.77 | 4.57 | 5.08 | 4.58 | 4.83 | 5.76 | 6.46 | 6.03 | 4.77 | 5.16 | 5.64 | 5.95 | 5.40 | 4.90 | 4.91 | 4.19 | 3.08 | 1.88 | 0.25 | -125 |
| 1 | 0.32 | 0.00 | 0.85 | 0.67 | 2.07 | 3.54 | 4.33 | 4.83 | 4.36 | 4.63 | 5.61 | 6.32 | 5.93 | 4.62 | 5.07 | 5.60 | 5.96 | 5.43 | 5.01 | 5.07 | 4.35 | 3.29 | 2.11 | 0.26 | -96 |
| 2 | 0.64 | 0.85 | 0.00 | 1.02 | 2.32 | 3.71 | 4.52 | 5.06 | 4.46 | 4.67 | 5.48 | 6.13 | 5.65 | 4.50 | 4.77 | 5.16 | 5.42 | 4.84 | 4.28 | 4.28 | 3.55 | 2.45 | 1.27 | 0.27 | -49 |
| 3 | 0.89 | 0.67 | 1.02 | 0.00 | 1.42 | 2.89 | 3.69 | 4.20 | 3.69 | 3.96 | 4.94 | 5.66 | 5.29 | 3.95 | 4.44 | 5.02 | 5.43 | 4.95 | 4.66 | 4.82 | 4.12 | 3.19 | 2.14 | 0.28 | -2 |
| 4 | 2.31 | 2.07 | 2.32 | 1.42 | 0.00 | 1.48 | 2.27 | 2.78 | 2.31 | 2.62 | 3.73 | 4.51 | 4.27 | 2.77 | 3.50 | 4.29 | 4.88 | 4.57 | 4.69 | 5.06 | 4.48 | 3.87 | 3.13 | 0.31 | 48 |
| 5 | 3.77 | 3.54 | 3.71 | 2.89 | 1.48 | 0.00 | 0.82 | 1.38 | 0.84 | 1.19 | 2.47 | 3.29 | 3.23 | 1.62 | 2.66 | 3.64 | 4.40 | 4.32 | 4.85 | 5.41 | 4.98 | 4.70 | 4.27 | 0.33 | 79 |
| 6 | 4.57 | 4.33 | 4.52 | 3.69 | 2.27 | 0.82 | 0.00 | 0.57 | 0.46 | 0.84 | 2.18 | 2.99 | 3.12 | 1.62 | 2.76 | 3.80 | 4.62 | 4.66 | 5.37 | 5.99 | 5.62 | 5.45 | 5.07 | 0.31 | 25 |
| 7 | 5.08 | 4.83 | 5.06 | 4.20 | 2.78 | 1.38 | 0.57 | 0.00 | 0.91 | 1.14 | 2.31 | 3.07 | 3.34 | 1.98 | 3.11 | 4.15 | 5.00 | 5.10 | 5.87 | 6.52 | 6.17 | 6.02 | 5.64 | 0.27 | 86 |
| 8 | 4.58 | 4.36 | 4.46 | 3.69 | 2.31 | 0.84 | 0.46 | 0.91 | 0.00 | 0.42 | 1.77 | 2.59 | 2.68 | 1.16 | 2.30 | 3.34 | 4.16 | 4.22 | 4.97 | 5.61 | 5.27 | 5.18 | 4.90 | 0.33 | 78 |
| 9 | 4.83 | 4.63 | 4.67 | 3.96 | 2.62 | 1.19 | 0.84 | 1.14 | 0.42 | 0.00 | 1.35 | 2.17 | 2.28 | 0.84 | 1.97 | 3.01 | 3.85 | 3.97 | 4.80 | 5.48 | 5.19 | 5.20 | 5.02 | 0.33 | 78 |
| 10 | 5.76 | 5.61 | 5.48 | 4.94 | 3.73 | 2.47 | 2.18 | 2.31 | 1.77 | 1.35 | 0.00 | 0.82 | 1.07 | 0.99 | 1.27 | 2.14 | 2.99 | 3.33 | 4.44 | 5.21 | 5.10 | 5.42 | 5.54 | 0.31 | 28 |
| 11 | 6.46 | 6.32 | 6.13 | 5.66 | 4.51 | 3.29 | 2.99 | 3.07 | 2.59 | 2.17 | 0.82 | 0.00 | 0.77 | 1.75 | 1.51 | 2.01 | 2.77 | 3.27 | 4.52 | 5.31 | 5.32 | 5.80 | 6.06 | 0.28 | -52 |
| 12 | 6.03 | 5.93 | 5.65 | 5.29 | 4.27 | 3.23 | 3.12 | 3.34 | 2.68 | 2.28 | 1.07 | 0.77 | 0.00 | 1.61 | 0.89 | 1.24 | 2.03 | 2.51 | 3.75 | 4.55 | 4.57 | 5.12 | 5.47 | 0.31 | 28 |
| 13 | 4.77 | 4.62 | 4.50 | 3.95 | 2.77 | 1.62 | 1.62 | 1.98 | 1.16 | 0.84 | 0.99 | 1.75 | 1.61 | 0.00 | 1.14 | 2.18 | 3.02 | 3.13 | 4.02 | 4.73 | 4.50 | 4.64 | 4.59 | 0.36 | 153 |
| 14 | 5.16 | 5.07 | 4.77 | 4.44 | 3.50 | 2.66 | 2.76 | 3.11 | 2.30 | 1.97 | 1.27 | 1.51 | 0.89 | 1.14 | 0.00 | 1.04 | 1.89 | 2.10 | 3.17 | 3.94 | 3.86 | 4.29 | 4.59 | 0.35 | 143 |
| 15 | 5.64 | 5.60 | 5.16 | 5.02 | 4.29 | 3.64 | 3.80 | 4.15 | 3.34 | 3.01 | 2.14 | 2.01 | 1.24 | 2.18 | 1.04 | 0.00 | 0.86 | 1.27 | 2.55 | 3.35 | 3.45 | 4.16 | 4.73 | 0.32 | 56 |
| 16 | 5.95 | 5.96 | 5.42 | 5.43 | 4.88 | 4.40 | 4.62 | 5.00 | 4.16 | 3.85 | 2.99 | 2.77 | 2.03 | 3.02 | 1.89 | 0.86 | 0.00 | 0.74 | 2.06 | 2.83 | 3.11 | 4.02 | 4.79 | 0.28 | -16 |
| 17 | 5.40 | 5.43 | 4.84 | 4.95 | 4.57 | 4.32 | 4.66 | 5.10 | 4.22 | 3.97 | 3.33 | 3.27 | 2.51 | 3.13 | 2.10 | 1.27 | 0.74 | 0.00 | 1.34 | 2.13 | 2.36 | 3.30 | 4.12 | 0.30 | 0 |
| 18 | 4.90 | 5.01 | 4.28 | 4.66 | 4.69 | 4.85 | 5.37 | 5.87 | 4.97 | 4.80 | 4.44 | 4.52 | 3.75 | 4.02 | 3.17 | 2.55 | 2.06 | 1.34 | 0.00 | 0.80 | 1.11 | 2.25 | 3.30 | 0.28 | -29 |
| 19 | 4.91 | 5.07 | 4.28 | 4.82 | 5.06 | 5.41 | 5.99 | 6.52 | 5.61 | 5.48 | 5.21 | 5.31 | 4.55 | 4.73 | 3.94 | 3.35 | 2.83 | 2.13 | 0.80 | 0.00 | 0.74 | 1.96 | 3.15 | 0.25 | -108 |
| 20 | 4.19 | 4.35 | 3.55 | 4.12 | 4.48 | 4.98 | 5.62 | 6.17 | 5.27 | 5.19 | 5.10 | 5.32 | 4.57 | 4.50 | 3.86 | 3.45 | 3.11 | 2.36 | 1.11 | 0.74 | 0.00 | 1.23 | 2.41 | 0.27 | -44 |
| 21 | 3.08 | 3.29 | 2.45 | 3.19 | 3.87 | 4.70 | 5.45 | 6.02 | 5.18 | 5.20 | 5.42 | 5.80 | 5.12 | 4.64 | 4.29 | 4.16 | 4.02 | 3.30 | 2.25 | 1.96 | 1.23 | 0.00 | 1.23 | 0.27 | -51 |
| 22 | 1.88 | 2.11 | 1.27 | 2.14 | 3.13 | 4.27 | 5.07 | 5.64 | 4.90 | 5.02 | 5.54 | 6.06 | 5.47 | 4.59 | 4.73 | 4.79 | 4.12 | 3.30 | 3.15 | 2.41 | 1.23 | 0.00 | | 0.27 | -58 |

Table 3. Calculation of W(u) for each node in the network
W(u) = (1/6) *(P1+P2+P3+P4+P5+P6)

| Node(u) | Deg (u) (P1) | CCI(u) (P2) | 1/ecc(u) (P3) | 1/MHD(u) (P4) | 1/MED(u) (P5) | NS(u) (P6) | W(u) |
|---|---|---|---|---|---|---|---|
| 0 | 4 | -86 | 0.14 | 0.27 | 0.25 | 275 | 32.28 |
| 1 | 5 | -25 | 0.17 | 0.31 | 0.26 | 325 | 50.96 |
| 2 | 3 | -51.5 | 0.14 | 0.26 | 0.27 | 175 | 21.20 |
| 3 | 5 | 8 | 0.17 | 0.29 | 0.28 | 400 | 68.96 |
| 4 | 3 | 2 | 0.14 | 0.27 | 0.31 | 300 | 50.95 |
| 5 | 3 | 34.5 | 0.17 | 0.29 | 0.33 | 200 | 39.72 |
| 6 | 3 | 5.5 | 0.14 | 0.28 | 0.31 | 175 | 30.71 |
| 7 | 2 | -81.5 | 0.14 | 0.25 | 0.27 | 100 | 3.53 |
| 8 | 2 | 32 | 0.14 | 0.28 | 0.33 | 125 | 26.63 |
| 9 | 3 | 44 | 0.17 | 0.29 | 0.33 | 200 | 41.30 |
| 10 | 3 | -7.5 | 0.17 | 0.27 | 0.31 | 250 | 41.04 |
| 11 | 5 | -9.5 | 0.17 | 0.30 | 0.28 | 375 | 61.88 |
| 12 | 3 | -7.5 | 0.14 | 0.27 | 0.31 | 250 | 41.04 |
| 13 | 5 | 99.5 | 0.17 | 0.31 | 0.36 | 300 | 67.56 |
| 14 | 3 | 91 | 0.17 | 0.30 | 0.35 | 175 | 44.97 |
| 15 | 2 | 18 | 0.17 | 0.28 | 0.32 | 75 | 15.96 |
| 16 | 4 | 34 | 0.17 | 0.32 | 0.28 | 225 | 43.96 |
| 17 | 1 | -26 | 0.14 | 0.26 | 0.30 | 100 | 12.62 |
| 18 | 4 | 38 | 0.17 | 0.33 | 0.28 | 275 | 52.96 |
| 19 | 3 | -37 | 0.17 | 0.29 | 0.25 | 125 | 15.29 |
| 20 | 1 | -86 | 0.14 | 0.23 | 0.27 | 50 | -5.73 |
| 21 | 3 | 19 | 0.17 | 0.33 | 0.27 | 150 | 28.80 |
| 22 | 2 | -8 | 0.17 | 0.31 | 0.27 | 125 | 19.96 |